\documentclass[amsmath,aps,showpacs,a4paper,10pt]{revtex4}

 \usepackage{epsf}
 \usepackage{graphicx}    

\begin{document}

 \newcommand{\be}[1]{\begin{equation}\label{#1}}
 \newcommand{\ee}{\end{equation}}
 \newcommand{\bea}{\begin{eqnarray}}
 \newcommand{\eea}{\end{eqnarray}}
 \def\disp{\displaystyle}

 \def\gsim{ \lower .75ex \hbox{$\sim$} \llap{\raise .27ex \hbox{$>$}} }
 \def\lsim{ \lower .75ex \hbox{$\sim$} \llap{\raise .27ex \hbox{$<$}} }

 \begin{titlepage}

 \begin{flushright}
 arXiv:1302.0643
 \end{flushright}

 \title{\Large \bf Cosmological Constraints
  on Variable Warm Dark Matter}

 \author{Hao~Wei\,}
 \email[\,email address:\ ]{haowei@bit.edu.cn}
 \affiliation{School of Physics, Beijing Institute
 of Technology, Beijing 100081, China}

 \author{Zu-Cheng~Chen\,}
 \affiliation{School of Physics, Beijing Institute
 of Technology, Beijing 100081, China}

 \author{Jing~Liu}
 \affiliation{School of Physics, Beijing Institute
 of Technology, Beijing 100081, China}

 \begin{abstract}\vspace{1cm}
 \centerline{\bf ABSTRACT}\vspace{2mm}
 Although $\Lambda$CDM model is very successful in many
 aspects, it has been seriously challenged. Recently, warm
 dark matter (WDM) remarkably rose as an alternative of cold
 dark matter (CDM). In the literature, many attempts have
 been made to determine the equation-of-state parameter (EoS)
 of WDM. However, in most of the previous works, it is usually
 assumed that the EoS of dark matter (DM) is constant (and
 usually the EoS of dark energy is also constant). Obviously,
 this assumption is fairly restrictive. It is more natural
 to assume a variable EoS for WDM (and dark energy). In the
 present work, we try to constrain the EoS of variable WDM
 with the current cosmological observations. We find that
 the best fits indicate WDM, while CDM is still consistent
 with the current observational data. However, $\Lambda$CDM
 is still better than WDM models from the viewpoint of
 goodness-of-fit. So, in order to distinguish WDM and CDM,
 the further observations on the small/galactic scale are
 required. On the other hand, in this work we also consider
 WDM whose EoS is constant, while the role of dark energy is
 played by various models. We find that the cosmological
 constraint on the constant EoS of WDM is fairly robust.
 \end{abstract}


 \maketitle

 \end{titlepage}

 \renewcommand{\baselinestretch}{1.1}


\section{Introduction}\label{sec1}

Nowadays, dark energy has become one of the most active fields
 in physics and astronomy, since the great discovery of the
 current accelerated expansion of our universe in
 1998~\cite{r1,r20}. On the other hand, dark matter (DM) was
 invoked to interpret the rotation curves of spiral galaxies
 for many years~\cite{r2}. Since it is well known that hot
 dark matter (HDM) cannot be competent for the cosmological
 structure formation, cold dark matter (CDM) has become the
 leading candidate. In fact, the well-known $\Lambda$CDM model
 has been established as the standard model in cosmology
 today~\cite{r1}.

Although $\Lambda$CDM model is very successful in many aspects,
 recently it has been seriously challenged. According to the
 brief reviews in e.g.~\cite{r3}, these serious challenges to
 $\Lambda$CDM model include, for example, (1)~$\Lambda$CDM
 predicts significantly smaller amplitude and scale of
 large-scale velocity flows than observations; (2) $\Lambda$CDM
 predicts fainter Type Ia supernova (SNIa) at high redshift
 $z$; (3) $\Lambda$CDM predicts more dwarf or irregular
 galaxies in voids than observed; (4) $\Lambda$CDM predicts
 shallow low concentration and density profiles of cluster
 haloes in contrast to observations; (5) $\Lambda$CDM predicts
 galaxy halo mass profiles with cuspy cores and low outer
 density while observations indicate a central core of
 constant density and a flattish high dark mass density outer
 profile; (6) $\Lambda$CDM predicts a smaller fraction of
 disk galaxies due to recent mergers expected to disrupt cold
 rotationally supported disks. Even when one replaces the
 cosmological constant $\Lambda$ with other (dynamical)
 dark energy candidates, these challenges still cannot be
 successfully addressed. In particular, the main source of the
 challenges on the small/galactic scale might be CDM. We refer
 to e.g.~\cite{r3} for details.

On the other hand, recently warm dark matter (WDM) remarkably
 rose as an alternative of CDM. The leading WDM candidates are
 the keV scale sterile neutrinos. In fact, the keV scale WDM
 is an intermediate case between the eV scale HDM and the GeV
 scale CDM. Unlike CDM which is challenged on the
 small/galactic scale (as mentioned above), it is claimed that
 WDM can successfully reproduce the astronomical observations
 over all the scales (from small/galactic to large/cosmological
 scales)~\cite{r4}. The key is the connection between the mass
 of DM particles and the free-streaming length $\ell_{\rm fs}$
 (structure smaller than $\ell_{\rm fs}$ will be erased). The
 eV scale HDM is too light and hence all structures below Mpc
 scale will be erased; the GeV scale CDM is too heavy and
 hence the structures below kpc scale cannot be erased
 (therefore CDM is challenged on the small/galactic scale);
 in between, the keV scale WDM works well~\cite{r4}. We refer
 to e.g.~\cite{r4} for a comprehensive review.

It is well known that the equation-of-state parameter (EoS)
 plays an important role in cosmology. In particular, the EoS
 of CDM and radiation/HDM are $0$ and $1/3$, respectively. In
 between, the EoS of WDM, $w_m$, satisfies $0\leq w_m\leq 1/3$
 obviously. Of course, the realistic value of $w_m$ should be
 determined from the astronomical observations. A non-zero
 $w_m$ indicates WDM, rather than CDM. In the literature, many
 attempts have been made to determine the EoS of DM. For
 example, in~\cite{r5}, assuming a constant $w_m$ and the role
 of dark energy is played by a cosmological constant $\Lambda$
 whose EoS is $-1$ exactly, it is found that
 $-1.50\times 10^{-6}<w_m<1.13\times 10^{-6}$ if DM produces
 no entropy, and $-8.78\times 10^{-3}<w_m<1.86\times 10^{-3}$
 if the adiabatic sound speed vanishes. Note that in~\cite{r5}
 the allowed range of the EoS of DM was relaxed and could be
 negative, but this is somewhat unnatural and hence it is not
 the case which will be investigated in the present work.
 Following the method proposed in~\cite{r6} which suggested to
 measure the EoS of DM by combining kinematic and gravitational
 lensing data, the authors of~\cite{r7} found that the EoS of
 DM is consistent with pressureless dark matter within the
 errors. The authors of~\cite{r8} considered the observational
 constraints on a cosmological model with variable EoS of
 matter and dark energy, namely, $w_m=1/[3(x^\alpha+1)]$ and
 $w_{de}=\bar{w}x^\alpha/(x^\alpha+1)$, where
 $x\equiv a/a_\ast$ with $a_\ast$ being a reference value of
 the scale factor $a$, and $\alpha>0$ is a constant model
 parameter. The model considered in~\cite{r8} was
 proposed to unify the radiation dominated phase and the dark
 energy dominated phase. The matter behaves as radiation when
 $x\ll 1$ and DM when $x\gg 1$. Note that this motivation is
 not for WDM, and its EoS $w_m$, $w_{de}$ are ad~hoc in some
 sense. The authors of~\cite{r9} considered the cosmological
 constraints on WDM whose EoS is a constant, while the EoS of
 dark energy is also a constant. They claimed that the
 cosmological data favor $w_m=0.006\pm 0.001$ (suggesting WDM),
 and $w_{de}=-1.11\pm 0.03$ (corresponding to phantom dark
 energy). The CDM whose EoS $w_m=0$ and the cosmological
 constant $\Lambda$ whose EoS $w_{de}=-1$ are disfavored
 beyond $3\sigma$ confidence level.

Note that in most of the previous works, it is usually assumed
 that the EoS of DM is constant (and usually the EoS of dark
 energy is also constant). Obviously, this assumption is fairly
 restrictive. It is more natural to assume a variable EoS for
 WDM (and dark energy). In fact, this is the main subject
 of the present work. We will try to constrain the EoS of
 variable WDM with the current cosmological observations. In
 Sec.~\ref{sec2}, we briefly introduce the observational data
 used in this work. In Sec.~\ref{sec3}, we consider the
 cosmological constraints on WDM whose EoS is variable. In
 fact, we adopt the familiar Chevallier-Polarski-Linder (CPL)
 parameterization for WDM, namely, $w_m=w_{m0}+w_{ma}(1-a)$.
 Unlike the somewhat ad hoc parameterization in~\cite{r8},
 noting that the Taylor series expansion of any function
 $F(x)$ is given by $F(x)=F(x_0)+F_1\,(x-x_0)+(F_2/\,2!)\,
 (x-x_0)^2+(F_3/\,3!)\,(x-x_0)^3+\dots$, the CPL
 parameterization for WDM can be regarded as the Taylor series
 expansion of $w_m$ with respect to the scale factor $a$ up to
 first order (linear expansion), and hence it is naturally
 motivated. In this section, we consider three cases, namely,
 the role of dark energy is played by a cosmological constant
 $\Lambda$, and dark energy described by constant EoS, CPL
 parameterized EoS, respectively. In Sec.~\ref{sec4}, noting
 that in most of the previous works constant EoS of both WDM
 and dark energy are assumed, here we also consider WDM whose
 EoS is constant, but dark energy is a cosmological constant
 $\Lambda$, or dark energy whose EoS are constant and CPL
 parameterized, respectively. We try to see whether the
 cosmological constraint on the constant EoS of WDM is robust
 for various types of dark energy, especially when the EoS of
 dark energy is variable. In Sec.~\ref{sec5}, some concluding
 remarks are given.


\section{Observational data}\label{sec2}

Recently, the Supernova Cosmology Project (SCP) Collaboration
 released the Union2.1 compilation which consists of 580 Type
 Ia supernovae (SNIa)~\cite{r10}. The Union2.1 compilation is
 the largest published and spectroscopically confirmed SNIa
 sample to date. The data points of the 580 Union2.1 SNIa
 compiled in~\cite{r10} are given in terms of the distance
 modulus $\mu_{obs}(z_i)$. On the other hand, the theoretical
 distance modulus is defined as
 \be{eq1}
 \mu_{th}(z_i)\equiv 5\log_{10}D_L(z_i)+\mu_0\,,
 \ee
 where $\mu_0\equiv 42.38-5\log_{10}h$ and $h$ is the Hubble
 constant $H_0$ in units of $100~{\rm km/s/Mpc}$, whereas
 \be{eq2}
 D_L(z)=(1+z)\int_0^z \frac{d\tilde{z}}{E(\tilde{z};{\bf p})}\,,
 \ee
 in which ${\bf p}$ denotes the model parameters; $z$ is the
 redshift; $E\equiv H/H_0$, in which $H\equiv\dot{a}/a$ is the
 Hubble parameter; a dot denotes the derivative with respect
 to cosmic time $t$; $a=(1+z)^{-1}$ is the scale factor (we
 have set $a_0=1$; the subscript ``0'' indicates the present
 value of corresponding quantity). Correspondingly, the
 $\chi^2$ from 580 Union2.1 SNIa is given by
 \be{eq3}
 \chi^2_{\mu}({\bf p})=\sum\limits_{i}\frac{\left[
 \mu_{obs}(z_i)-\mu_{th}(z_i)\right]^2}{\sigma^2(z_i)}\,,
 \ee
 where $\sigma$ is the corresponding $1\sigma$ error. The
 parameter $\mu_0$ is a nuisance parameter but it
 is independent of the data points. One can perform a uniform
 marginalization over $\mu_0$. However, there is an alternative
 way. Following~\cite{r11,r12}, the minimization with respect
 to $\mu_0$ can be made by expanding the $\chi^2_{\mu}$ of
 Eq.~(\ref{eq3}) with respect to $\mu_0$ as
 \be{eq4}
 \chi^2_{\mu}({\bf p})=\tilde{A}-2\mu_0\tilde{B}+\mu_0^2\tilde{C}\,,
 \ee
 where
 $$\tilde{A}({\bf p})=\sum\limits_{i}\frac{\left[
 \mu_{obs}(z_i)-\mu_{th}(z_i;\mu_0=0,{\bf p})\right]^2}
 {\sigma_{\mu_{obs}}^2(z_i)}\,,$$
 $$\tilde{B}({\bf p})=\sum\limits_{i}\frac{\mu_{obs}(z_i)
 -\mu_{th}(z_i;\mu_0=0,{\bf p})}{\sigma_{\mu_{obs}}^2(z_i)}\,,
 ~~~~~~~~~~~
 \tilde{C}=\sum\limits_{i}\frac{1}{\sigma_{\mu_{obs}}^2(z_i)}\,.$$
 Eq.~(\ref{eq4}) has a minimum for $\mu_0=\tilde{B}/\tilde{C}$
 at
 \be{eq5}
 \tilde{\chi}^2_{\mu}({\bf p})=\tilde{A}({\bf p})
 -\frac{\tilde{B}({\bf p})^2}{\tilde{C}}\,.
 \ee
 Since $\chi^2_{\mu,\,min}=\tilde{\chi}^2_{\mu,\,min}$ (up to
 a constant) obviously, we can instead minimize
 $\tilde{\chi}^2_{\mu}$ which is independent of $\mu_0$.
 In addition to SNIa, the other useful observations include the
 cosmic microwave background (CMB) anisotropy~\cite{r13} and
 the large-scale structure (LSS)~\cite{r14}. However, using
 the full data of CMB and LSS to perform a global fitting
 consumes a large amount of computation time and power. As an
 alternative, one can instead use the shift parameter $R$ from
 CMB, and the distance parameter $A$ of the measurement of
 the baryon acoustic oscillation (BAO) peak in the distribution
 of SDSS luminous red galaxies. In the literature, the shift
 parameter $R$ and the distance parameter $A$ have been used
 extensively. It is argued in e.g.~\cite{r15} that they are
 model-independent and contain the main information of the
 observations of CMB and BAO, respectively. As is well known,
 the shift parameter $R$ of CMB is defined by~\cite{r15,r16}
 \be{eq6}
 R\equiv\Omega_{m0}^{1/2}\int_0^{z_\ast}
 \frac{d\tilde{z}}{E(\tilde{z})}\,,
 \ee
 where the redshift of recombination $z_\ast=1091.3$ which was
 determined by the Wilkinson Microwave Anisotropy Probe
 7-year (WMAP7) data~\cite{r13}, and
 $\Omega_{m0}\equiv 8\pi G\rho_{m0}/(3H_0^2)$ is the present
 fractional density of matter. The shift parameter $R$ relates
 the angular diameter distance to the last scattering surface,
 the comoving size of the sound horizon at $z_\ast$ and the
 angular scale of the first acoustic peak in CMB power spectrum
 of temperature fluctuations~\cite{r15,r16}. The value of $R$
 has been determined to be $1.725\pm 0.018$ from the WMAP7
 data~\cite{r13}. On the other hand, the distance parameter $A$
 of the measurement of the BAO peak in the distribution of SDSS
 luminous red galaxies~\cite{r14} is given by
 \be{eq7}
 A\equiv\Omega_{m0}^{1/2}E(z_b)^{-1/3}\left[\frac{1}{z_b}
 \int_0^{z_b}\frac{d\tilde{z}}{E(\tilde{z})}\right]^{2/3},
 \ee
 where $z_b=0.35$. In~\cite{r17}, the value of $A$ has been
 determined to be $0.469\,(n_s/0.98)^{-0.35}\pm 0.017$. Here
 the scalar spectral index $n_s$ is taken to be $0.963$,
 which comes from the WMAP7 data~\cite{r13}. So, the total
 $\chi^2$ is given by
 \be{eq8}
 \chi^2=\tilde{\chi}^2_{\mu}+\chi^2_{CMB}+\chi^2_{BAO}\,,
 \ee
 where $\tilde{\chi}^2_{\mu}$ is given in Eq.~(\ref{eq5}),
 $\chi^2_{CMB}=(R-R_{obs})^2/\sigma_R^2$ and
 $\chi^2_{BAO}=(A-A_{obs})^2/\sigma_A^2$. The best-fit model
 parameters are determined by minimizing the total $\chi^2$.
 As in~\cite{r18,r19,r20}, the $68.3\%$ confidence level is
 determined by $\Delta\chi^2\equiv\chi^2-\chi^2_{min}\leq 1.0$,
 $2.3$, $3.53$, $4.72$ and $5.89$ for $n_p=1$, $2$, $3$, $4$
 and $5$ respectively, where $n_p$ is the number of free model
 parameters. Similarly, the $95.4\%$ confidence level is
 determined by $\Delta\chi^2\equiv\chi^2-\chi^2_{min}\leq 4.0$,
 $6.18$, $8.02$, $9.72$ and $11.31$ for $n_p=1$, $2$, $3$, $4$
 and $5$, respectively.


\section{Cosmological constraints on variable WDM}\label{sec3}

In this section, we consider the cosmological constraints
 on WDM whose EoS is variable. Throughout this work, we
 consider a flat Friedmann-Robertson-Walker (FRW) universe
 which contains dark energy and WDM (we assume that radiation
 and the baryon component can be ignored). Here, we adopt the
 familiar CPL parameterization for WDM, namely, its EoS is given by
 \be{eq9}
 w_m=w_{m0}+w_{ma}(1-a)\,,
 \ee
 where $w_{m0}$ and $w_{ma}$ are both constants. Unlike the
 somewhat ad hoc parameterization used in~\cite{r8}, as
 mentioned in Sec.~\ref{sec1}, the CPL parameterization
 for WDM can be regarded as the Taylor series expansion~of
 $w_m$ with respect to the scale factor $a$ up to first order
 (linear expansion), and~hence~it~is~naturally~motivated.
 To ensure $0\leq w_m\leq 1/3$ in the whole
 history ($0\leq a\leq 1$) as mentioned in Sec.~\ref{sec1},
 we require~that
 \be{eq10}
 0\leq w_{m0}\leq 1/3\,,~~~~~~~0\leq w_{m0}+w_{ma}\leq 1/3\,.
 \ee
 In this section, we consider three cases, namely, the
 role of dark energy is played by a cosmological constant
 $\Lambda$, and dark energy described by constant EoS, CPL
 parameterized EoS, respectively.


 \begin{center}
 \begin{figure}[tbhp]
 \centering
 \includegraphics[width=0.5\textwidth]{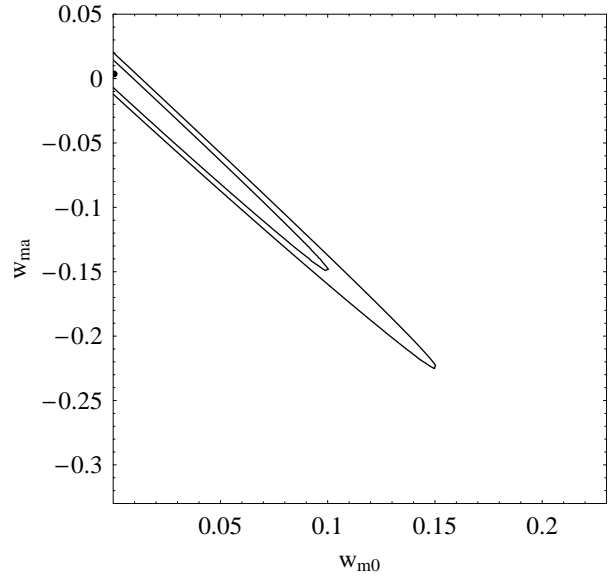}
 \caption{\label{fig1}
 The $68.3\%$ and $95.4\%$ confidence level contours in the
 $w_{m0}-w_{ma}$ plane for the $\Lambda$VWDM model. The
 best-fit parameters are also indicated by a black solid
 point.}
 \end{figure}
 \end{center}


\vspace{5mm} 


 \begin{center}
 \begin{figure}[bbhp]
 \centering
 \includegraphics[width=0.5\textwidth]{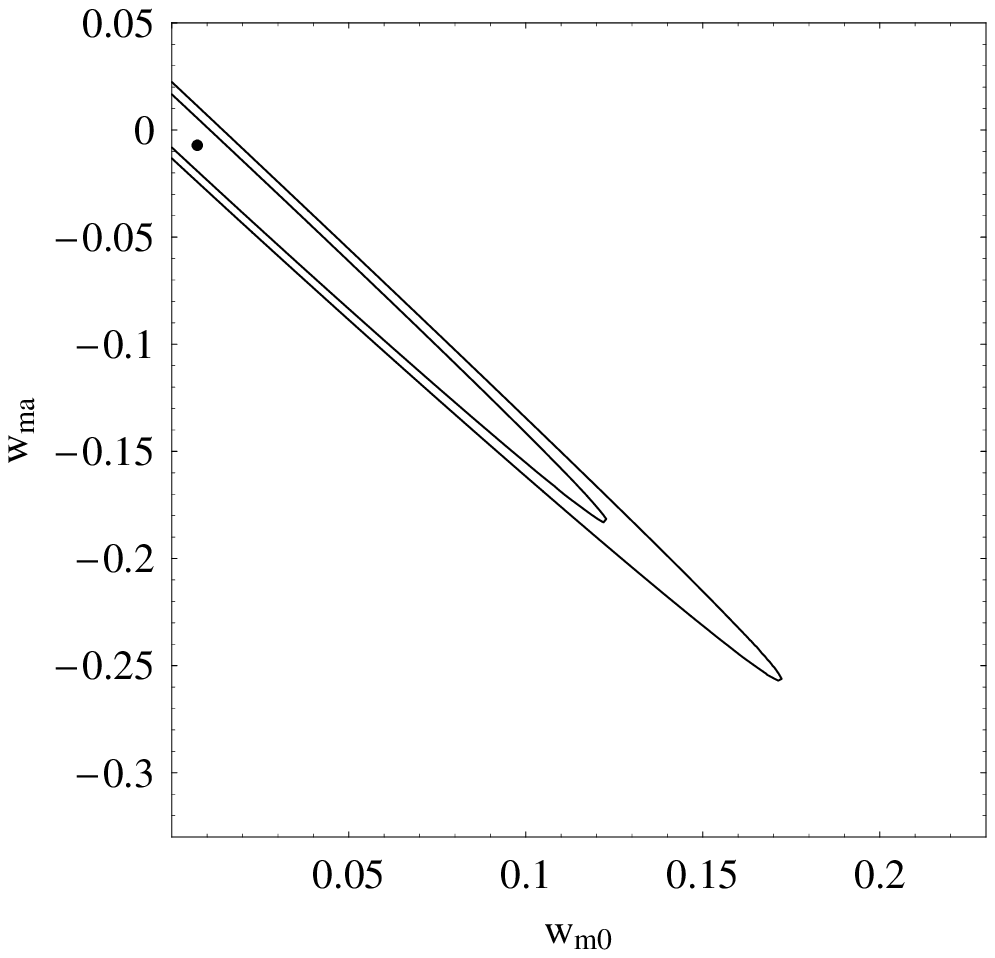}
 \caption{\label{fig2}
 The $68.3\%$ and $95.4\%$ confidence level contours in the
 $w_{m0}-w_{ma}$ plane for the XVWDM model. The best-fit
 parameters are also indicated by a black solid point.}
 \end{figure}
 \end{center}



 \begin{center}
 \begin{figure}[tbhp]
 \centering
 \includegraphics[width=1.0\textwidth]{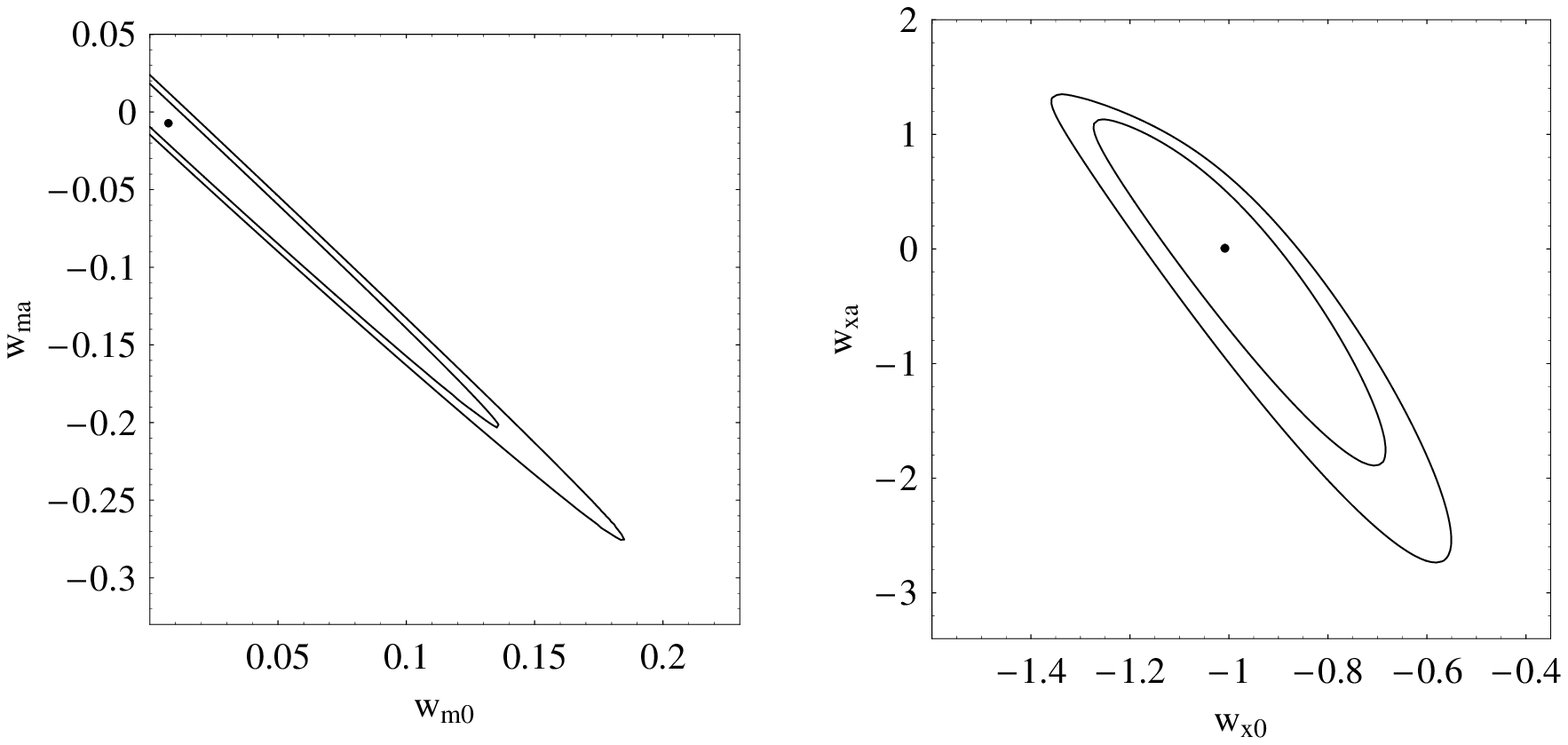}
 \caption{\label{fig3}
 The $68.3\%$ and $95.4\%$ confidence level contours in the
 $w_{m0}-w_{ma}$ plane and the $w_{\rm x0}-w_{\rm xa}$ plane
 for the CVWDM model. The best-fit parameters are also
 indicated by a black solid
 point.\\} 
 \end{figure}
 \end{center}


At first, we consider the case in which the role of dark energy
 is played by a cosmological constant $\Lambda$. We call it
 $\Lambda$VWDM model. As is well known, the corresponding $E(z)$ is
 given by~\cite{r19,r21,r22}
 \be{eq11}
 E(z)=\left[\Omega_{m0}(1+z)^{3(1+w_{m0}+w_{ma})}
 \exp\left(-\frac{3w_{ma} z}{1+z}\right)+
 \left(1-\Omega_{m0}\right)\right]^{1/2}.
 \ee
 There are 3 free parameters in this model. By minimizing the
 corresponding total $\chi^2$ in Eq.~(\ref{eq8}), we find the
 best-fit parameters $\Omega_{m0}=0.2774$, $w_{m0}=0.0$ and
 $w_{ma}=0.0036$, while $\chi^2_{min}=562.227$. In Fig.~\ref{fig1},
 we present the corresponding $68.3\%$ and $95.4\%$ confidence
 level contours in the $w_{m0}-w_{ma}$ plane for
 the $\Lambda$VWDM model. It is easy to see that the best fit
 indicates WDM, while CDM is still consistent with the
 current observational data. Note that for the best fit,
 $w_m=0$ at $z=0$, namely it is CDM today while it is WDM in
 the past ($a<1$).

Next, we consider the case in which the EoS of dark energy
 $w_{\rm x}$ is a constant. We call it XVWDM model. In this
 case, the corresponding $E(z)$ is given by~\cite{r19,r21,r22}
 \be{eq12}
 E(z)=\left[\Omega_{m0}(1+z)^{3(1+w_{m0}+w_{ma})}
 \exp\left(-\frac{3w_{ma} z}{1+z}\right)+
 \left(1-\Omega_{m0}\right)(1+z)^{3(1+w_{\rm x})}\right]^{1/2}.
 \ee
 There are 4 free parameters in this model. By minimizing the
 corresponding total $\chi^2$ in Eq.~(\ref{eq8}), we find the
 best-fit parameters $\Omega_{m0}=0.2773$, $w_{m0}=0.0072$,
 $w_{ma}=-0.0071$, and $w_{\rm x}=-1.0072$, while
 $\chi^2_{min}=562.225$. In Fig.~\ref{fig2}, we present the
 corresponding $68.3\%$ and $95.4\%$ confidence level contours
 in the $w_{m0}-w_{ma}$ plane for the XVWDM model. Again, it
 is easy to see that the best fit indicates WDM, while CDM is
 still consistent with the current observational data. Note
 that for the best fit, it is always WDM in the whole history
 ($0\leq a\leq 1$), and $w_m=w_{m0}=0.0072$ today. Comparing
 Figs.~\ref{fig1} and~\ref{fig2}, we can easily find that the
 confidence level contours become larger.

Finally, we consider the case in which the EoS of dark energy
 $w_{\rm x}$ is also variable. Here, we adopt CPL
 parameterization for dark energy, namely
 \be{eq13}
 w_{\rm x}=w_{\rm x0}+w_{\rm xa}(1-a)\,,
 \ee
 where $w_{\rm x0}$ and $w_{\rm xa}$ are both constants. We
 call it CVWDM model. In this case, the corresponding $E(z)$
 is given by~\cite{r19,r21,r22}
 \bea
 &&E(z)=\left[\Omega_{m0}(1+z)^{3(1+w_{m0}+w_{ma})}
 \exp\left(-\frac{3w_{ma} z}{1+z}\right)\right.\nonumber\\
 &&\hspace{3cm}\left.+\left(1-\Omega_{m0}\right)
 (1+z)^{3(1+w_{\rm x0}+w_{\rm xa})}\exp\left(
 -\frac{3w_{\rm xa} z}{1+z}\right)\right]^{1/2}.\label{eq14}
 \eea
 There are 5 free parameters in this model. By minimizing the
 corresponding total $\chi^2$ in Eq.~(\ref{eq8}), we find the
 best-fit parameters $\Omega_{m0}=0.2773$, $w_{m0}=0.0072$,
 $w_{ma}=-0.0072$, $w_{\rm x0}=-1.0081$, and
 $w_{\rm xa}=0.0062$, while $\chi^2_{min}=562.225$. In
 Fig.~\ref{fig3}, we present the corresponding $68.3\%$ and
 $95.4\%$ confidence level contours in the $w_{m0}-w_{ma}$
 plane and the $w_{\rm x0}-w_{\rm xa}$ plane for
 the CVWDM model. From Fig.~\ref{fig3}, we see
 that $\Lambda$CDM is still consistent with the current
 observational data. Note that for the best fit, it is always
 WDM in the whole history ($0<a\leq 1$), and
 $w_m=w_{m0}=0.0072$ today.


 \begin{center}
 \begin{figure}[tbhp]
 \centering
 \includegraphics[width=0.5\textwidth]{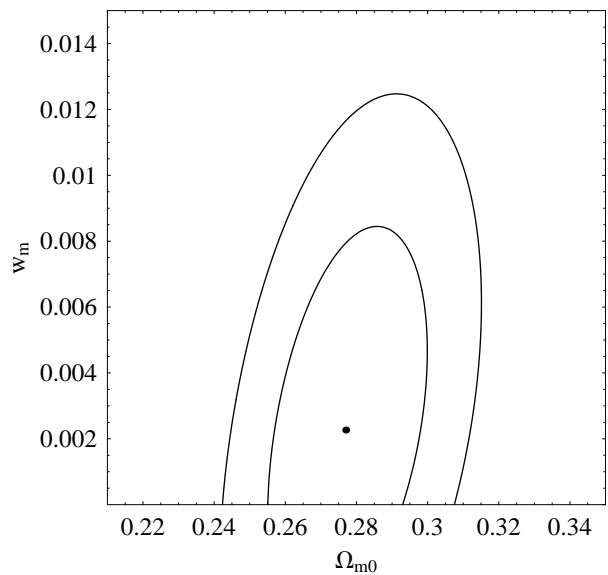}
 \caption{\label{fig4}
 The $68.3\%$ and $95.4\%$ confidence level contours in the
 $\Omega_{m0}-w_m$ plane for the $\Lambda$WDM model. The
 best-fit parameters are also indicated by a black solid
 point.}
 \end{figure}
 \end{center}


\vspace{-10mm} 


\section{Cosmological constraints on WDM with constant EoS}\label{sec4}

Noting that in most of the previous works constant EoS of both
 WDM and dark energy are assumed, here we also consider WDM
 whose EoS $w_m$ is constant, but dark energy is a cosmological
 constant $\Lambda$, or dark energy whose EoS are constant and
 CPL parameterized, respectively. We try to see whether the
 cosmological constraint on the constant EoS of WDM is robust
 for various types of dark energy, especially when the EoS of
 dark energy is variable.

We firstly consider the case with a cosmological constant
 $\Lambda$. We call it $\Lambda$WDM model. Note that the EoS
 of WDM should satisfy $0\leq w_m\leq 1/3$ as mentioned in
 Sec.~\ref{sec1}. As is well known, the corresponding $E(z)$ is
 given by~\cite{r19,r21,r22}
 \be{eq15}
 E(z)=\left[\Omega_{m0}(1+z)^{3(1+w_m)}+\left(1-\Omega_{m0}
 \right)\right]^{1/2}.
 \ee
 There are 2 free parameters in this model. By minimizing the
 corresponding total $\chi^2$ in Eq.~(\ref{eq8}), we find the
 best-fit parameters $\Omega_{m0}=0.2770$ and $w_m=0.0023$,
 while $\chi^2_{min}=562.228$. In Fig.~\ref{fig4}, we present
 the corresponding $68.3\%$ and $95.4\%$ confidence level
 contours in the $\Omega_{m0}-w_m$ plane for the $\Lambda$WDM
 model. It is easy to see that the best fit indicates WDM,
 while CDM is still consistent with the current observational data.

\begin{center}
 \begin{figure}[htbp]
 \centering
 \includegraphics[width=0.5\textwidth]{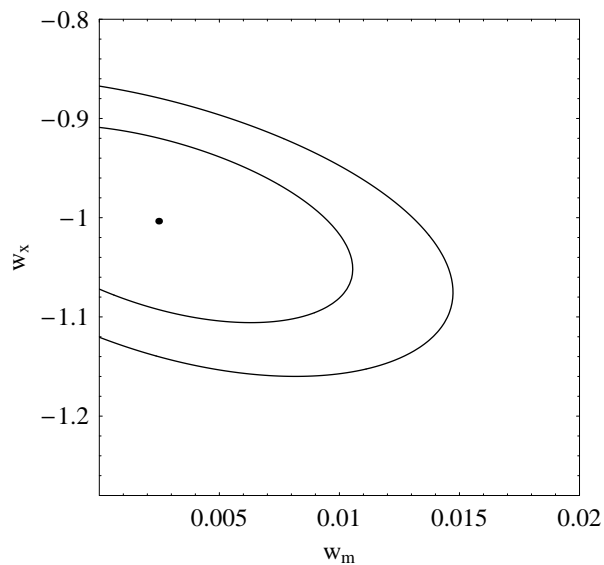}
 \caption{\label{fig5}
 The $68.3\%$ and $95.4\%$ confidence level contours in the
 $w_m-w_{\rm x}$ plane for the XWDM model. The best-fit
 parameters are also indicated by a black solid point.}
 \end{figure}
 \end{center}

Then, we turn to the case in which the EoS of dark energy
 $w_{\rm x}$ is a constant. We call it XWDM model. In this
 case, the corresponding $E(z)$ is given by~\cite{r19,r21,r22}
 \be{eq16}
 E(z)=\left[\Omega_{m0}(1+z)^{3(1+w_m)}+\left(1-
 \Omega_{m0}\right)(1+z)^{3(1+w_{\rm x})}\right]^{1/2}.
 \ee
 There are 3 free parameters in this model. By minimizing the
 corresponding total $\chi^2$ in Eq.~(\ref{eq8}), we find the
 best-fit parameters $\Omega_{m0}=0.2776$, $w_m=0.0025$, and
 $w_{\rm x}=-1.0035$, while $\chi^2_{min}=562.225$. In
 Fig.~\ref{fig5}, we present the corresponding $68.3\%$ and
 $95.4\%$ confidence level contours in the $w_m-w_{\rm x}$
 plane for the XWDM model. Again, the best fit indicates WDM,
 while CDM is still consistent with the current observational
 data.


\vspace{-3mm} 

\begin{center}
 \begin{figure}[tbhp]
 \centering
 \includegraphics[width=1.0\textwidth]{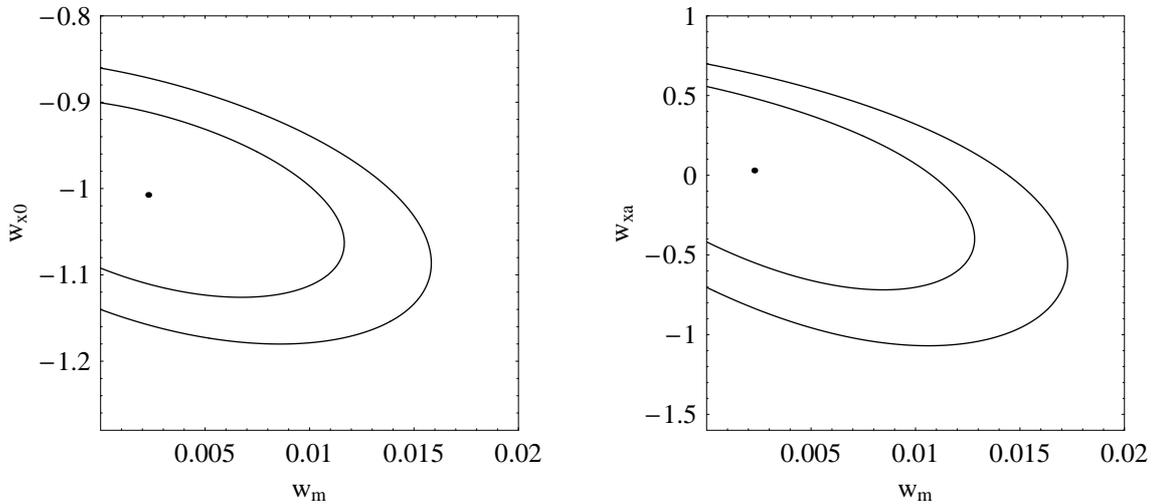}
 \caption{\label{fig6}
 The $68.3\%$ and $95.4\%$ confidence level contours in the
 $w_m-w_{\rm x0}$ plane and the $w_m-w_{\rm xa}$ plane
 for the CWDM model. The best-fit parameters are also
 indicated by a black solid point.}
 \end{figure}
 \end{center}

Finally, we consider the case with CPL parameterized dark
 energy, namely, the EoS of dark energy is given
 by Eq.~(\ref{eq13}). We call it CWDM model. In this case,
 the corresponding $E(z)$ is given by~\cite{r19,r21,r22}
 \be{eq17}
 E(z)=\left[\Omega_{m0}(1+z)^{3(1+w_m)}+\left(1-\Omega_{m0}
 \right)(1+z)^{3(1+w_{\rm x0}+w_{\rm xa})}
 \exp\left(-\frac{3w_{\rm xa} z}{1+z}\right)\right]^{1/2}.
 \ee
 There are 4 free parameters in this model. By minimizing the
 corresponding total $\chi^2$ in Eq.~(\ref{eq8}), we find the
 best-fit parameters $\Omega_{m0}=0.2773$, $w_m=0.0023$,
 $w_{\rm x0}=-1.0074$, and $w_{\rm xa}=0.0299$,
 while $\chi^2_{min}=562.224$. In Fig.~\ref{fig6}, we present
 the corresponding $68.3\%$ and $95.4\%$ confidence level
 contours in the $w_m-w_{\rm x0}$ plane and the
 $w_m-w_{\rm xa}$ plane for the CWDM model. From
 Fig.~\ref{fig6}, we see that $\Lambda$CDM is still consistent
 with the current observational data.

Comparing these three cases, it is easy to find that their
 best-fit $w_m$ are almost the same value. This indicates that
 the cosmological constraint on the constant EoS of WDM is
 fairly robust, namely it is insensitive to the dark energy models.


\section{Concluding remarks}\label{sec5}

Although $\Lambda$CDM model is very successful in many
 aspects, it has been seriously challenged. Recently, warm
 dark matter (WDM) remarkably rose as an alternative of cold
 dark matter (CDM). In the literature, many attempts have
 been made to determine the equation-of-state parameter (EoS)
 of WDM. However, in most of the previous works, it is usually
 assumed that the EoS of dark matter (DM) is constant (and
 usually the EoS of dark energy is also constant). Obviously,
 this assumption is fairly restrictive. It is more natural
 to assume a variable EoS for WDM (and dark energy). In the
 present work, we try to constrain the EoS of variable WDM
 with the current cosmological observations. We find that
 the best fits indicate WDM, while CDM is still consistent
 with the current observational data. On the other hand, in
 this work we also consider WDM whose EoS is constant, while
 the role of dark energy is played by various models. We find
 that the cosmological constraint on the constant EoS of WDM
 is fairly robust.

As mentioned above, in the six cases considered in this work,
 all the best fits indicate WDM, while CDM is still consistent
 with the current observational data. So, it is worthwhile to
 compare these WDM models with $\Lambda$CDM. To this end, we
 also fit $\Lambda$CDM model to the same observational data,
 and find the best-fit $\Omega_{m0}=0.2736$, while
 $\chi^2_{min}=562.546$. Following~\cite{r19,r20}, here we
 adopt three criterions used extensively in the literature,
 namely $\chi^2_{min}/dof$, Bayesian Information Criterion
 (BIC) and Akaike Information Criterion (AIC). Note that the
 degree of freedom $dof=N-k$, whereas $N$ and $k$ are the
 number of data points and the number of free model parameters,
 respectively. The BIC is defined by~\cite{r23}
 \be{eq18}
 {\rm BIC}=-2\ln{\cal L}_{max}+k\ln N\,,
 \ee
 where ${\cal L}_{max}$ is the maximum likelihood. In the
 Gaussian cases, $\chi^2_{min}=-2\ln{\cal L}_{max}$. Thus,
 the difference in BIC between any two models is given by
 \be{eq19}
 \Delta{\rm BIC}=\Delta\chi^2_{min}+\Delta k \ln N\,.
 \ee
 On the other hand, the AIC is defined by~\cite{r24}
 \be{eq20}
 {\rm AIC}=-2\ln{\cal L}_{max}+2k\,.
 \ee
 Correspondingly, the difference in AIC between any two models
 reads
 \be{eq21}
 \Delta{\rm AIC}=\Delta\chi^2_{min}+2\Delta k\,.
 \ee
 In Table~\ref{tab1}, we present $\chi^2_{min}/dof$, $\Delta$BIC and
 $\Delta$AIC for $\Lambda$CDM and all the six WDM models considered
 in this work. Notice that $\Lambda$CDM has been chosen to be
 the fiducial model when we calculate $\Delta$BIC and $\Delta$AIC.
 From Table~\ref{tab1}, we see that the rank of models is coincident
 in all the three criterions ($\chi^2_{min}/dof$, BIC and AIC).
 Obviously, $\Lambda$CDM model is the best one. In summary, although
 the best fits to the cosmological observations (SNIa, CMB and BAO)
 indicate WDM, we cannot say WDM is favored, since $\Lambda$CDM
 is still better than WDM models from the viewpoint of
 $\chi^2_{min}/dof$, BIC and AIC. So, in order to distinguish
 WDM and CDM, the further observations on the small/galactic
 scale are required.


 \begin{table}[tbp]
 \begin{center}
 \begin{tabular}{llllllll} \hline \hline \\[-3.5mm]
 Model & $\Lambda$CDM & $\Lambda$VWDM & XVWDM
 & CVWDM & $\Lambda$WDM & XWDM & CWDM
 \\[1.2mm] \hline \\[-3.5mm]
 $\chi^2_{min}$ & 562.546 & 562.227 & 562.225 & 562.225
 & 562.228 & 562.225 & 562.224 \\
 $k$ & 1 & 3 & 4 & 5 & 2 & 3 & 4 \\
 $\chi^2_{min}/dof$~~~ & 0.968238~~~ & 0.971031~~~
 & 0.972708~~~ & 0.974393~~~ & 0.969359~~~ & 0.971028~~~
 & 0.972706 \\
 $\Delta$BIC & 0 & 12.4139 & 18.7784 & 25.1449 & 6.04847
 & 12.4119 & 18.7774 \\
 $\Delta$AIC & 0 & 3.681 & 5.679 & 7.679 & 1.682
 & 3.679 & 5.678 \\
 Rank & 1 & 4 & 6 & 7 & 2 & 3 & 5
 \\[1.2mm] \hline\hline
 \end{tabular}
 \end{center}
 \caption{\label{tab1} Comparing all the six WDM models
 with $\Lambda$CDM.}
 \end{table}



\section*{ACKNOWLEDGEMENTS}
We thank the anonymous referee for quite useful comments and
 suggestions, which helped us to improve this work. We are
 grateful to Professors Rong-Gen~Cai, Shuang~Nan~Zhang,
 Xiao-Jun~Bi for helpful discussions. We also thank Minzi~Feng,
 as well as Long-Fei~Wang, Xiao-Jiao~Guo and Xiao-Peng~Yan,
 for kind help and discussions. This work was supported in
 part by NSFC under Grants No.~11175016 and No.~10905005,
 as well as NCET under Grant No.~NCET-11-0790, and the
 Fundamental Research Fund of Beijing Institute of Technology.

\renewcommand{\baselinestretch}{1.1}


\end{document}